\documentclass[12pt]{iopart}
\usepackage[latin1]{inputenc}
\usepackage[T1]{fontenc}
\usepackage[english]{babel}
\usepackage{graphicx}
\begin{document}
\title[Exact solutions of the Hyperextended Scalar Tensor theory with potential...]{Exact solutions of the Hyperextended Scalar Tensor theory with potential in the Bianchi type I model.}
\author{Stéphane Fay \footnote{Steph.Fay@Wanadoo.fr}\\
14 rue de l'Etoile\\
75017 Paris\\
France
}
\begin{abstract}
The Hyperextended Scalar Tensor theory with a potential is defined by three free functions: the gravitational function, the Brans-Dicke coupling function and the potential. Starting from the expression of the 3-volume and the potential as function of the proper time, we determine the exact solutions of this theory. We study two important cases corresponding to power and exponential laws for the 3-volume and the potential. 
\\
\\
\\
Published in Classical and Quantum Gravity copyright 2001 IOP Publishing Ltd\\
Classical and Quantum Gravity, Vol 18, 1, 2001.\\
http://www.iop.org
\end{abstract}
\maketitle
\section{Introduction} \label{s1}
In this work, we study the Hyperextended Scalar Tensor theory (HST) with a potential for the Bianchi type I model. We determine, by help of quadratures, the exact solutions of this theory as function of the 3-volume and the potential. We then study the case for which these 2 functions are power or exponential laws of the proper time.

The simplest scalar tensor theories is the Brans-Dicke theory studied in the sixties by Brans and Dicke \cite{BraDic61}. One of their goals was to integrate the Mach's principle in a gravitational relativistic theory. Since the eighties, other justifications have been given to take into account scalar fields in gravitational theories. They are issued from inflation and particles physics theories, i.e. the unification models. Their low energy limit can be described by scalar tensor theories. As instance, it is the case for supergravity theory or higher dimensional theories. Large classes of scalar tensor theories belong to the HST \cite{TorVuc96}. Its Lagrangian is written as this of the Brans-Dicke one but the coupling constant is replaced by a function of the scalar field and the gravitational function , $\phi$ in the Brans-Dicke theory, by any function $G^{-1}(\phi)$. In this paper, we will also consider a potential $U(\phi)$ which is predicted by particles physics for the early epochs. Moreover, recent studies on the type I supernovae \cite{Per99,Rie98} could confirm the presence of a cosmological constant, which would be the remainder of the potential for late time epochs. We will not consider other type of matter as perfect fluids, thus assuming a Universe dominated by the scalar field. Such phases for the Universe are relevant: near the singularity, perfect fluid are often negligible \cite{Gas99}. Furthermore, it is sometimes considered that scalar fields could be responsible for a large part of the dark matter which could be the dominant matter of our Universe.

Lets justify the geometrical framework of this paper. It is well known that large scale structures we observe could not exist if the Universe has always respected the cosmological principle. Thus it seems reasonable to consider other models such as the homogeneous ones: these are the Bianchi models. Some of them have the interesting property to isotropize toward an FLRW model: the Bianchi type $V$ model can approach the open FLRW one, the Bianchi type $IX$ model can tend toward the closed FLRW one and the Bianchi type $I$ model toward the flat FLRW one. Recent observations from Boomerang \cite{Lan00} seem favour closed models. However, from the point of view of inflationary models, the flat model is the most studied and is usually preferred to other ones. Hence, it is difficult today to decide what is the best model to describe our Universe and we will choose to study the Bianchi type $I$ one.

Now, we give a more accurate description of the functions characterising the HST. When a potential is present, it is defined by three free functions: the function $G$ playing the role of a variable gravitational function, the Brans-Dicke coupling function describing the coupling between the scalar field and the metric, and last the potential $U$. To find exact solutions, most of times one choose $G(\phi)$, $\omega(\phi)$ and $U(\phi)$ and determine the form of the metric functions. However, other methods exist to achieve this goal which have been mainly applied to the FLRW models. As instance, in the fine turning potential method \cite{Bar78, Che97, MaaTayRou95}, one first choose the form of the metric functions and then look for the form of the potential. In this paper, we will assume that the forms of the potential and the 3-volume of the Universe are known functions of the proper time. Despite interesting works to determine the form of the potential from the observations \cite{Sta98, EspPol00}, there is no method today to predict it for the HST. The second quantity is related to the isotropic part $e^\Omega$ of the metric\footnote{We have then that the 3-volume is equal to $e^{3\Omega}$.} or the scale factor of the FLRW models. We will use theoretical considerations to choose its form as a function of the proper time. From these two quantities, we can get the exact forms of the gravitational function $G(t)$ and the anisotropic part of the metric. Moreover, if we choose a form for $\phi(t)$, we obtain $\omega(t)$ and then the theory and the solution of the field equations are fully determined.

Our motivations are the following:
\begin{itemize}
\item To find the exact solutions of the field equations when we know the isotropic part of the metric and the potential as functions of the proper time. This is a mathematical motivation corresponding to an extension of the fine turning potential.
\item To study the dynamical properties of the Universe (isotropisation, inflation...) for special forms of $e^{\Omega}$ and $U$, i.e. power law and exponential law of the proper time in this paper. This deserves physical motivation since most of this results could be extendable to any function $e^{\Omega}$ and $U$ asymptotically tending toward these special forms. Thus the scalar tensor theories whose the isotropic part and the potential can be asymptotically written as power series of $t$ or exponential of $t$ will be concerned by these results.
\end{itemize}

This paper is organised as follows: in the section \ref{s2} we write the field equations and give their exact solutions. In section \ref{s3}, we look for the properties of the models defined by $(e^\Omega=t^m,U=t^n)$ and $(e^\Omega=e^{mt},U=e^{nt})$. We conclude in section \ref{s4}.
\section{Exact solution of the field equations of the HST with potential in the Bianchi type $I$ model} \label{s2}
We use the following form of the metric:
\begin{equation} \label{1}
ds^{2}=-dt^{2}+e^{2\alpha} (\omega^1)^2+e^{2\beta}(\omega^2)^2+e^{2\gamma}(\omega^3)^2
\end{equation}
The $\omega_i$ are the 1-forms specifying the Bianchi type $I$ model. We introduce the parameterisation:
\begin{eqnarray*}
\alpha &=&\Omega+\beta_+\\
\beta &=&\Omega+\beta_-\\
\gamma &=&\Omega-\beta_+-\beta_-\\
\end{eqnarray*}
It is similar to this of Misner \cite{Mis69}. The function $\Omega$ stands for the isotropic part of the metric and the functions $\beta_\pm$ describe the anisotropic part. The isotropic part is linked to the 3-volume $V$ of the Universe by the relation $V=e^{3\Omega}$. The action of the HST is written:
\begin{equation} \label{2}
S=\int (G^{-1}R-\frac{\omega}{\phi}\phi_{,\mu}\phi^{,\mu}-U)\sqrt{g}dx^4
\end{equation}
$\phi$ is the scalar field, $U$ the potential, $\omega$ the Brans-Dicke coupling function and $G$ the gravitational function. Lets justify the study of this action.
Since in this paper we will have no need to assume any relation between the scalar field and the Brans-Dicke coupling function, we could use the action of the Generalised Scalar Tensor theory (GST) which has the same form as (\ref{2}) but with $G^{-1}=\phi$. However we do not want to impose any relation between $G$ and the scalar field since $G^{-1}=\phi$ is not the only form of the gravitation function in the literature. As instance String theory at low energy is defined by $G^{-1}=e^{\phi}$ and important studies have been made with gravitational function writing as $G^{-1}=\phi^2+constant$. Hence, although we have no need for this in this paper, we will consider a general form for $G$. Lets underline that some HST theories can not be cast into a GST when the function $G$ is not invertible although this change of variable is then singular and could be an indication for mathematical inconsistency\footnote{I thank one of the referees for having clarified this point.}.\\
The action (\ref{2}) could also be equivalent to the General Relativity plus a minimally coupled scalar field if we redefine the metric functions as $\bar g_{\mu\nu}=G^{-1}g_{\mu\nu}$\footnote{Lets note that some results for the General Relativity with a minimally coupled scalar field can be get from these of the HST by putting $G^{-1}=1$. But for obvious reasons it is not so simple to  get results for the HST from these of the General Relativity with a minimally coupled scalar field.}. We get then the so-called Einstein frame and the metric (\ref{1}) is the so-called Brans-Dicke frame. We have chosen to work with the last metric since the results we would get in the Einstein frame would not have been equivalent to these of the Brans-Dicke frame: as shown, as instance, in \cite{Fay00A}, and contrary to what happens when we do a "simple" scalar field redefinition, the conditions for the isotropisation in both frames are not always the same. The same conclusion arises for the presence or not of inflation. Thus, the results get in the Brans-Dicke frame for the HST will not be equivalent to these found in the Einstein frame or/and for General Relativity with minimally coupled scalar field, it is rather a generalisation. Moreover, the Brans-Dicke frame is generally assumed to be the physical one, although this point can be subject to discussion. One could also ask why we have not first studied the Einstein frame and then extended our results to the Brans-Dicke one. However, to proceed we would have to integrate $G^{-1}(\bar t)$, which is not always workable.\\

We get the field equations by varying the action with respect to the metric functions. In the $\tau$ time defined by $dt=Vd\tau$ we obtain:
\begin{equation} \label{3}
\alpha^{,,}+\alpha^, G G^{-1,}+1/2G G^{-1,,}-1/2GV^2U=0
\end{equation}
and similar equations for $\beta$ and $\gamma$. The prime stands for the derivative with respect to $\tau$. For the constraint, we get 
\begin{equation} \label{4a}
\alpha^,\beta^,+\alpha^,\gamma^,+\beta^,\gamma^,+GG^{-1,}V^,V^{-1}-1/2UGV^{2}-1/2\omega G\phi^{,2}\phi^{-1}=0
\end{equation}
By adding the three spatial components, we find a differential equation for the 3-volume:
\begin{equation} \label{4}
V^{,,}V^{-1}G^{-1,}-V^{,2}V^{-1}G^{-1}+V^,V^{-1}G^{-1,}+3/2G^{-1,,}-3/2UV^2=0
\end{equation}
If we use equation (\ref{4}) to express $UV^2$ and introduce this quantity in (\ref{3}), we have for $\beta_\pm$:
\begin{equation} \label{5}
\beta_{\pm}=\beta_{\pm 0}\int G e^{-3\Omega} dt+\beta_{\pm 1}
\end{equation}
$\beta_{\pm 0}$ and $\beta_{\pm 1}$ are integration constants. Thus, the Universe isotropize when $t\rightarrow \infty$ if $\int G e^{-3\Omega}dt$ tends toward a constant. Now, we want to evaluate the gravitational function $G$ depending on $\Omega$ and $U$. We find with help of (\ref{4}):
\begin{equation}\label{6}
G^{-1}=e^{-2\Omega}\left[\int{\frac{\int{Ue^{3\Omega} dt}+G_0}{e^{\Omega}}dt+G_1}\right]
\end{equation}
$G_0$ and $G_1$ are constants. We can make two remarks:
\begin{itemize}
\item We have completely determined $G(t)$ and $\beta_\pm(t)$ as functions of $\Omega(t)$ and $U(t)$. The solutions of the spatial field equations are independent on the form of the scalar field and the Brans-Dicke coupling function since they depend only on the gravitational coupling function which is expressed as a function of the proper time and not of the scalar field.
\item Moreover, we can write $G^{-1}$ as:
\begin{equation}
G^{-1}=g_1(\Omega)+g_2(\Omega,U)
\end{equation}
Then, writing $U=\sum_{n} U_n$, we see that $G^{-1}(\Omega ,\sum_n U_n)=g_1(\Omega)+\sum_n g_2(\Omega,U_n)$. Thus, from the solution of the field equations  for n potentials, we should be able to determine the solution for their sum. As instance, if we know then for a potential writing as $t^n$, we will be able to deduce the solution for any potential writing as a power law series.
\end{itemize}
Now, let's express the Brans-Dicke coupling function as function of $\Omega$, $U$ and $\phi$. Using the constraint equation and (\ref{5}), we get:
\begin{equation} \label{5a}
\omega=2G^{-1}\dot{\phi}^{-2}\phi\left[3\dot{\Omega}^{2}-G^2 e^{-6\Omega}(\beta_{+0}^2+\beta_{-0}^2+\beta_{+0}\beta_{-0})+3G\dot{G^{-1}}\dot{\Omega}-1/2GU\right]
\end{equation}
The Brans-Dicke coupling function is then fully defined by $\Omega(t)$, $U(t)$ and $\phi(t)$. It exists the same type of linearity relation between $\omega$ and $U$ as for $G$. We have:
\begin{equation}
\omega(\phi,\Omega,\sum_nU_n)=\omega_1(\phi,\Omega)+\sum_n\omega_2(\phi,\Omega,U_n)
\end{equation}
In the next section, we study two classes of models for which the isotropic part of the metric and the potential are power or exponential laws of the proper time. For the clarity of the discussion, we will assume that their solutions are defined in $t\rightarrow +\infty$, which represents the late times epoch.
\section{Application} \label{s3}
\subsection{Power laws} \label{s31}
We choose power law forms for the isotropic part of the metric and the potential:
\begin{eqnarray} \label{7}
&e^\Omega=t^m&\\
&U=U_0t^n&\nonumber\\\nonumber
\end{eqnarray}
$U_0$ is a constant. When the Universe isotropize, the metric functions tends toward $e^\Omega$ which can be then compared to the scale factor of the FLRW models. In the flat isotropic models, the scale factor often takes power law forms as for General Relativity with perfect fluid. It is also an important form for the inflation, which received the name of polynomial inflation. In a general way, the association of the forms (\ref{7}) is physically meaningful for many theories studied in the FLRW models. As instance, such forms for the scale factor and the potential have been found in \cite{Fei00} where a superpotential is considered. This is also asymptotically the case in \cite{AbrCraMim94} where conformal scalar field cosmologies are examined and in general for any forms of $e^\Omega$ and $U$ which can be asymptotically developed as power law series. Thus, the results which follow could apply to large class of scalar tensor theories. Additional reasons will be given in the next section.
\begin{enumerate}
\item \textbf{\emph{Gravitational function}}\\
From (\ref{6}), we get:
\begin{equation} \label{8a}
G^{-1}=C_1 t^{2+n} + C_2 t^{-2m} + C_3 t^{-3m+1}
\end{equation}
$C_i$ are integration constants. From (\ref{8a}), we see that we can not choose $m$ and $n$ such that $G$ be constant unlike asymptotically. Thus, General Relativity does not belong to the class of theories defined by these forms of $\Omega$ and $U$. When the Universe isotropizes, it will be in expansion if $m>0$ and will undergo inflation if $m>1$. The potential will tend naturally toward zero if $n<0$. From (\ref{5}) we deduce that isotropisation will happen at late times if $m>1$ or $3m+n>-1$. In the first case, the Universe will be necessarily inflationary for this period. In the last case, inflation will go with isotropisation if $n<-4$. Thus, we can get an isotropic Universe without inflation. Consequently, if $m>1$ or $3m+n>-1$, the Universe isotropises and the power law $t^m$ represents a late times attractor for the metric functions.\\
Power laws of the proper time for the gravitational function play an important role toward the literature. Milne, in the thirties, studied the case $G=t$ and Dirac, in the framework of the "Large Number hypothesis", proposed $G=t^{-1}$ \cite{Dir37}. More recently, in \cite{Bar97} a study of the Newtonian cosmologies with polynomial laws for $G$ and perfect fluid ($p=(\gamma-1)\rho$) in the isotropic models is made. It is shown that for $G=t^p$, inflation is present when $3\gamma>2$ and do not depends on the variation of $G$. In this work, a condition to get asymptotically inflation is $m>1$. In this case, $G\rightarrow t^{-(2+n)}$ if $n>-4$. Then, it shows that inflation in the class of models we are studying, is also asymptotically independent on the variation of the gravitational function as in \cite{Bar97} if the potential is larger than $t^{-4}$.
\item \textbf{\emph{Applications}}\\
In this part, we examine several known types of Universes which are late times attractor when isotropisation arises.
\begin{itemize}
\item Coasting Universe\\
We choose $m=1$. Then, the Universe isotropizes at late times and tends toward a coasting Universe, i.e. the metric functions tend toward $t$ if $n>-4$. We calculate the exact solutions of the field equations. The anisotropic part of the metric is written:
\begin{equation}
\beta_\pm=-\beta_{\pm 0}\frac{\ln \left[(C_2+C_3)t^{-4-n}+C_1\right]}{(4+n)(C_2+C_3)}+\beta_{\pm 1}
\end{equation}
Coasting Universe has been previously studied in \cite{PimDia98}. In this paper, a Brans-Dicke model with a perfect fluid and a power law potential in a FLRW model was considered. For open, closed or flat models, they found linear expansion of the scale factor with  a potential decreasing inversely with the square of time. An important characteristic of an isotropic coasting cosmologies is that the age of the Universe is not in conflict with the observations. So, the age problem is absent for this type of dynamical behaviour for the metric functions.
\item Cosmological constant\\
We examine the case $m=1/2$ and $n=0$. The potential is then a cosmological constant. At late times, the Universe isotropizes and the metric functions tend toward $t^{1/2}$ which is the form of the scale factor in an FLRW model for General Relativity when Universe is radiation dominated. This theory could thus build a bridge between an anisotropic Universe dominated by the scalar field and a flat relativistic and isotropic Universe dominated by the radiation. The gravitational function will behave asymptotically as $t^{-2}$ and then will tend to vanish at late times. The exact solution for the metric functions can be found. The anisotropic part of the metric is written as:
\begin{equation}
\beta_\pm=\beta_{\pm 0}\sum_{i=1}^6 \frac{\ln (\sqrt{t}+s_i)}{C_3+6C_1s_i^5}+\beta_{\pm 1}
\end{equation}
The $s_i$ are the  $i^{th}$ roots of the equation $C_2+C_3 s+C_1 s^6=0$. A cosmological constant is equivalent to consider an equation of state for the stiff fluid ($p=-\rho$). This situation has been studied in \cite{Bar97} where $G=t^p$. In this last paper it has been noticed that the asymptotical behaviour of the scale factor was determined by the value of $p$, the power of the gravitational function. Here, for the class of theories defined by (\ref{7}), whatever $m>0$, i.e. an asymptotically isotropic and expanding Universe, the gravitational function always behaves as $t^{-2}$ at late times in presence of a cosmological function. The behaviour of $G$ is thus independent on the value of $m$.\\
\item Gravitational constant\\
Another interesting case corresponds to $m=1/2$ and $n=-2$. The anisotropic part of the metric functions is written:
\begin{equation}
\beta_\pm=\beta_{\pm 0}4\arctan(\frac{C_3+2C_1\sqrt{t}}{(4C_1C_2-C_3^2)^{1/2}})(4C_1C_2-C_3^2)^{-1/2}+\beta_{\pm 1}
\end{equation}
Then, the potential tends to vanish at late times. The gravitational function tends asymptotically toward the constant $C_1$. One more times, this theory connects an anisotropic Universe to an isotropic one behaving dynamically as if radiation was present, but here the gravitational function tends toward a constant.  In a general manner, when $m>1/3$, the case $n=-2$ is the only one giving birth to an asymptotically non-vanishing gravitational constant. Since, recent observations suggest that our present Universe could undergo inflation, which means $m>1$, this remark underline the importance of a potential behaving like $t^{-2}$ at late times if we assume a gravitational constant for this epoch and a power law behaviour for the scale factor. This type of potential has been studied in \cite{CheZhu99, ZhuChe98} and particularly in \cite{PimDia98} where it arises naturally when one use a scalar field behaving as a power law type of the proper time.
\item Static Universe\\
For $m=0$ and $n>-1$, the Universe will isotropize toward a static behaviour at late times. If moreover we require that the potential be decreasing, we need $n\in\left[-1,0\right]$. The anisotropic part of the metric takes the form of a hypergeometric function multiplied by $t$.  Static phases for the Universe are interesting since they can help to solve the age problem and the problem of the large-scale structures formation.\\
\end{itemize}
\end{enumerate}
\subsection{Exponential laws} \label{s32}
\begin{figure}[h]
\begin{center}
\includegraphics{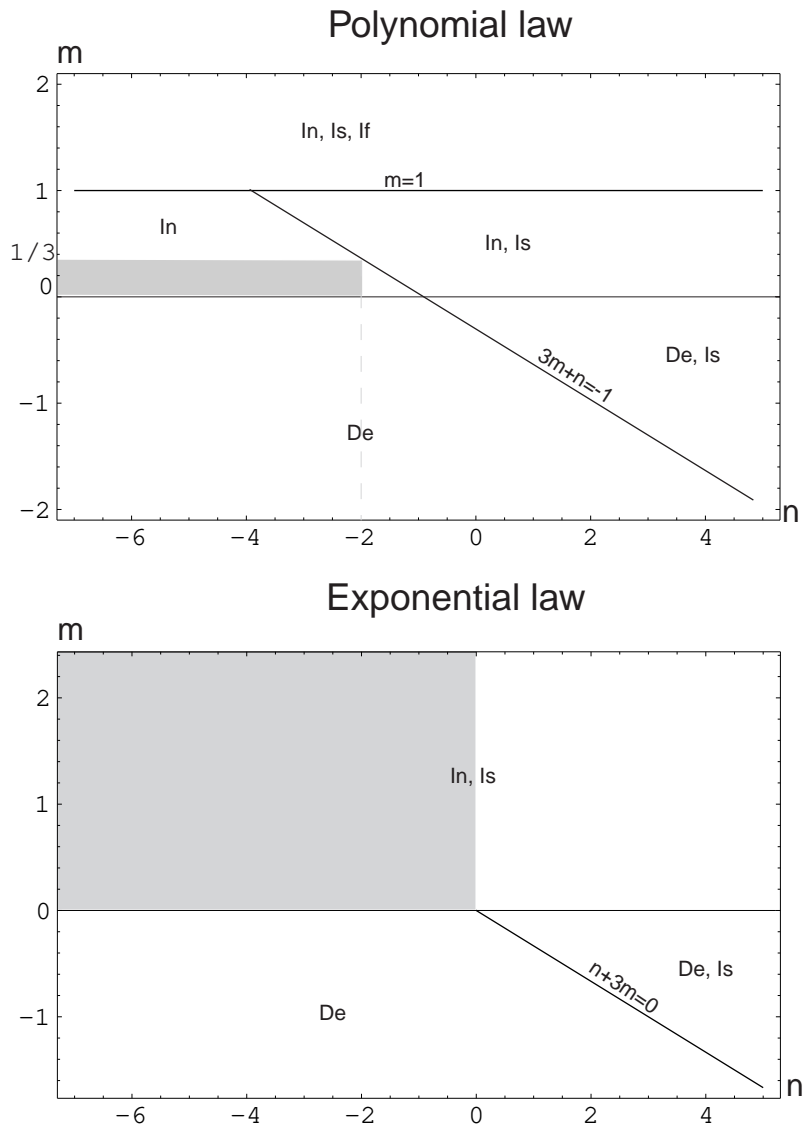}
\end{center}
\caption{These two figures summarise the asymptotical behaviours of the HST in the $(m,n)$ plane when the potential and the isotropic parts of the metric are respectively power or exponential laws of the proper times. We have annoted "De", "In", "Is" and "If" the regions of the plane where the Universe is respectively decreasing, increasing, isotropic and inflationary at late times epochs. Gray regions represent the regions of the plane for which the gravitational function  diverge at late times.}
\label{fig1}
\end{figure}

We choose an exponential law for the isotropic part of the metric and the potential:
\begin{eqnarray} \label{7a}
&e^\Omega=e^{mt}&\\
&U=U_0e^{nt}&\nonumber\\\nonumber
\end{eqnarray}
$U_0$ is a constant. When the Universe isotropizes and undergoes expansion, we get a De-Sitter like behaviour for its dynamics and thus exponential inflation. This justifies the importance of this case which can also be considered as a limiting case of the polynomial inflation with $m\rightarrow +\infty$. Moreover, in FLRW models, the association of exponential laws for the scale factor and the potential is recovered in \cite{Fei00} where a superpotential is used and in \cite{AbrCraMim94} where conformal scalar field cosmologies are considered.
\begin{enumerate}
\item \textbf{\emph{gravitational function}}\\
The gravitational function is written:
\begin{equation} \label{8}
G^{-1}=C_1 e^{nt} + C_2 e^{-3mt} + C_3 e^{-2mt}
\end{equation}
The $C_i$ are integration constant. The Universe will isotropize at late times if $m>0$, which means it is then expanding, or/and $n>-3m$. In this last case, asymptotically contracting Universe implies that the potential diverge. When the Universe isotropizes it tends toward a De-Sitter model. Consequently, when $m>0$ and/or $n>-3m$, a De-Sitter Universe is a late times attractor for the class of theories defined by (\ref{7a}).  This result can be compared, for the Bianchi type I model, to Wald results \cite{Wal83} which claims that, in the case of General Relativity with a scalar field and a cosmological constant, all the Bianchi models (except contracting Bianchi type IX) initially in expansion approach the isotropic De-Sitter solution.\\
The exact solution of the field equations can be found whatever $m$ and $n$. We get for the anisotropic part of the metric:
\begin{equation} \label{9}
\beta_\pm=\beta_{\pm 0}(mt-\ln \mbox{[}1+C_3e^{mt}(C_2+C_1e^{(3m+n)t})^{-1}\mbox{]})\mbox{[}m(C_2+C_1e^{(3m+n)t})\mbox{]}^{-1}+\beta_{\pm 1}
\end{equation}
If we choose the early times at $t\rightarrow -\infty$, the functions $\beta_\pm$ tend toward linear law of the proper time or constant. This means that at early times the metric functions tend toward exponential laws of the proper time which can be compared to an anisotropic De Sitter Universe.\\
The only asymptotical behaviour for the metric functions which could be common between the case of this subsection and the previous one is an asymptotical static Universe. A necessary condition is then $m=0$. Then, we see from (\ref{9}) that the Universe can not isotropize and thus, asymptotically static Universe is not possible for the class of theories defined by (\ref{7a}). This last result excludes that General Relativity with a scalar field and a cosmological constant, defined by $m=0$ and $n=0$, isotropize at late time with a constant scale factor. This is in accordance with Wald results.
\end{enumerate}
The late times behaviours of the classes of theories described in subsections \ref{s31} and \ref{s32} are represented on figure \ref{fig1}.\\
\\
To our knowledge the results get in this last section are new and most of then can be extended to any functions $e^\Omega$ and $U$ tending asymptotically toward the forms examined above. In \cite{Fay00A}, the same type of applications have been made in the Einstein frame. It was shown that the Universe tends toward an isotropic De-Sitter model when the potential tends toward a constant and reciprocally. In the present paper, we can see that such behaviour also arises when the potential diverges. In the same way, it was shown that the Universe isotropizes when its isotropic part tends toward a power law behaviour of the proper time if the scalar field is defined when the metric functions diverge but we had not get conditions on the exponent of the power law representing $e^\Omega$. Moreover in subsection \ref{s31} we have also shown that Universe can isotropize toward a static model which is not possible in the Einstein frame. This underlines that HST is not dynamically equivalent to General Relativity with a scalar field and that new dynamical behaviours can be found.
\section{Conclusion} \label{s4}
In this work, we have determined, with help of quadrature, the solution of the field equations of the HST with a potential in the Bianchi type $I$ model when we know the form of the potential and the isotropic part of the metric as some functions of the proper time. The first result we get is that the Universe isotropize when the integral of $Ge^{-3\Omega}$ tends asymptotically toward a constant. We had already obtained it in \cite{Fay00A} by help of Hamiltonian formalism. Physically, it means that the 3-volume of the Universe have to grow faster than the gravitational function. It is in accordance with our present Universe which is expanding with a probably constant gravitational function. The Brans-Dicke coupling function can be evaluated as a function of the proper time and finally of the scalar field if it is an invertible function of $t$. However we have not study any particular form of $\omega$ since the dynamical properties of the Universe does not depend on it.\\

We have looked for the exact solutions of two classes of theories respectively defined by power and exponential laws of the proper time for $e^\Omega$ and $U$. They lead to isotropic Universe with power or exponential inflation and are linked, among others, to the presence of superpotential or conformal scalar field cosmologies. Of course, the forms we have chosen for $e^\Omega$ and $U$ are particular ones. However most of the results we obtained should stay true for any theory whose the isotropic part of the metric and the potential asymptotically behave like these described in section \ref{s3}. Particularly power laws of $t$ are very interesting since from a mathematical point of view, any solutions which can be developed as power law series can be approximated in this way. Thus our results and the assumptions that the Universe be isotropic and undergoes inflation at late times could constraint any scalar tensor theories whose anisotropic part of the metric and potential can be developed as power series of the proper time or as exponential of $t$. Lets note also that from a physical point of view, power laws of the proper time are good approximations for the behaviour of the 3-volume at late times, when the solutions of the field equations approach FLRW ones, or at early times when the singularity is described by Kasnerian behaviour.\\

When the potential and the isotropic part of the metric are written as functions of power of $t$ (respectively $m$ and $n$), the gravitational function is the sum of three powers of the proper time. Then, the Universe isotropizes when $m>1$ or $3m+n>-1$. In these two cases, an isotropic Universe with a power law for the metric functions represents the late times attractor. In the first case, the Universe will undergo inflation. In the second case, the presence of inflation at late times will imply $n<-4$. The opposite is not true.\\
We can not found the exact form of the anisotropic part of the metric for any values of $m$ and $n$. However some interesting cases can be studied. The first one corresponds to an asymptotical coasting universe for which $m=1$. Then, at late times the dynamical behaviour of the metric functions when universe isotropizes, i.e. for $n>-4$, is a linear law of $t$. In the FLRW case, this theory does not suffer of the age problem. A second case corresponds to a Universe with a cosmological constant ($n=0$) which tends toward a power law of times with the same form as the solution for the flat radiation dominated Universe in the isotropic case ($m=1/2$). For these two values of $n$ and $m$, the Universe will always isotropize at late times. This theory is thus able to build a bridge between an anisotropic Bianchi type $I$ Universe with a cosmological constant and an isotropic one behaving dynamically like a flat isotropic radiation dominated Universe. In a general way, when we consider a cosmological constant and a power law for the isotropic part of the metric , if $m>0$, the Universe isotropize at late times and the gravitational function behaves like $t^{-2}$ and vanishes. Moreover, if instead of a cosmological constant we choose a potential behaving like $t^{-2}$, the gravitational function will tend toward a constant instead of vanishing which is a physically interesting situation since it is what we observe for $G$. Remark that, whatever $m$ and $n$, the only way to get asymptotically a non-vanishing gravitational constant with an expanding Universe is to choose $m=1/3$ and $n\leq -2$ or $n=-2$ and $m\geq 1/3$. Hence, if we want to get at late times an inflationary Universe with a gravitation constant, we have to choose a potential behaving as $t^{-2}$. Note that the potential will then vanish at late times and will diverge at early times ($t=0$), thus solving the cosmological constant problem. The last case we study is this of an asymptotical isotropic and static Universe ($m=0$). It will be a late times attractor, i.e. the Universe will always tend toward an isotropic and static Universe, if $n>-1$. This type of theory could help to solve age and large scale structures formation problems.\\
When the potential and the isotropic part of the metric are written as functions of exponential of $t$ (respectively $m$ and $n$), $G^{-1}$ is then the sum of three exponentials of $t$. The Universe will isotropize at late times if $m>0$ or/and $n>-3m$. Under these conditions for $m$ and $n$, a De-Sitter Universe is a late times attractor. In the first case, this always give birth to an expanding Universe. In the second case, if the potential asymptotically vanishes at late times as it could be the case for our present Universe, it can be contracting or expanding Universe.\\
It is possible to calculate the exact solutions of the metric functions, i.e. the anisotropic part of the metric, for any $m$ and $n$. Thus, we have remarked that this class of models can not isotropize asymptotically toward a static Universe ($m=0$). However, in the neighbourhood of the singularity that we choose in $t\rightarrow -\infty$, all the metric functions tend toward an exponential of the scalar field giving birth to the counterpart of a De Sitter model for the anisotropic case, i.e. the metric functions behave as exponentials of $t$ with different exponents.\\

The two classes of theories we present in this paper have large regions of the parameters plane $(m,n)$ for which the Universe is able to isotropize and be in expansion with a vanishing potential  (In this case, for exponential laws of $U$ and $e^{\Omega}$, the gravitational function always diverge. This is different when we consider power laws.). Such types of theories could solve the cosmological constant problem since their potentials decrease naturally when time increases. These regions of the plane $(m,n)$ are shown in the figure \ref{fig1}.

This work can be extended in several ways. First, since $G^{-1}(\Omega ,\sum_n U_n)=g_1(\Omega)+\sum_n g_2(\Omega ,U_n)$ and $\omega(\phi ,\Omega ,\sum_n U_n)=\omega_1(\phi,\Omega)+\sum_n \omega_2(\phi$ $,\Omega ,U_n)$, we are able from the dynamical properties of simple classes of theories defined by $\Omega(t)$ and $U(t)$ to deduce dynamical properties of more evoluted classes of theories defined by sum of these functions. Secondly, other types of physically interesting potentials can be studied such as these which tend toward a constant in an oscillating way as for instance $U=\sin t/t+U_0$. Finally, It would be interesting to extend this work to take into account the presence of a perfect fluid as matter field for the Universe. It will be the subject of future works.

\section*{References}
\bibliographystyle{unsrt}

\end{document}